\documentclass[11pt,a4paper]{article}
\pdfoutput=1
\usepackage{jcappub}

\usepackage[utf8]{inputenc}
\usepackage{amsmath,amsfonts,amssymb}
\usepackage{graphicx}
\usepackage{color}
\usepackage{subcaption}
\usepackage{float}
\usepackage{orcidlink}

\newcommand{\td}{{\rm d}}

\title{Primordial black holes from strong first-order phase transitions}

\author[1]{Marek Lewicki\,\orcidlink{0000-0002-8378-0107}\,,}
\author[1]{Piotr Toczek\,\orcidlink{0000-0003-4904-3063}\,,}
\author[2,3,4]{and Ville Vaskonen\,\orcidlink{0000-0003-0003-2259}\,}

\affiliation[1]{Institute of Theoretical Physics, Faculty of Physics, University of Warsaw,\\ ul. Pasteura 5, 02-093 Warsaw, Poland}
\affiliation[2]{Dipartimento di Fisica e Astronomia, Universit\`a degli Studi di Padova, Via Marzolo 8, 35131 Padova, Italy}
\affiliation[3]{Istituto Nazionale di Fisica Nucleare, Sezione di Padova, Via Marzolo 8, 35131 Padova, Italy}
\affiliation[4]{National Institute of Chemical Physics and Biophysics, R\"avala 10, Tallinn, Estonia}

\emailAdd{marek.lewicki@fuw.edu.pl}
\emailAdd{piotr.toczek@fuw.edu.pl}
\emailAdd{ville.vaskonen@pd.infn.it}

\abstract{We study the formation of primordial black holes (PBHs) in strongly supercooled first-order phase transitions. The mechanism is based on the presence of remnants dominated by the false vacuum that scale slower with the expansion of the Universe than their surroundings where this energy was already converted into radiation. We compute the PBH formation from these remnants including the contribution from the false vacuum and the bubble walls, by estimating the collapse using the hoop conjecture and by considering both regions collapsing immediately when entering the horizon and sub-horizon regions that collapse as their compactness grows. We show that for exponential bubble nucleation rate, $\Gamma \propto e^{\beta t}$, the primordial black hole formation implies $\beta/H \gtrsim 3.8$, where $H$ denotes the Hubble rate, if the potential energy of the false vacuum is $\Delta V \lesssim (10^{12} {\rm GeV})^4$, as otherwise a too large abundance of long-lived PBHs forms. The observed dark matter abundance can be formed in asteroid mass PBHs if $\beta/H \simeq 3.8$ and $10^5 {\rm GeV} \lesssim \Delta V^{1/4} \lesssim 10^8 {\rm GeV}$. Finally, we consider also the effect of the second order correction to the exponential nucleation rate showing that the PBH abundance is mainly determined by the average radius of the true vacuum bubbles.}

\notoc

\begin{document}

\maketitle

\section{Introduction} 

Various processes in the early universe may have led to the formation of black holes (BHs). The cosmological and astrophysical role of these primordial black holes (PBHs) can be very significant. In particular, PBHs in the asteroid mass range can explain all dark matter (DM)~\cite{Carr:2020gox}. Even if PBHs were so light that they evaporated quickly after their formation~\cite{Hawking:1974rv, Hawking:1975vcx}, they may have affected DM phenomenology~\cite{Fujita:2014hha, Allahverdi:2017sks, Lennon:2017tqq, Hooper:2019gtx, Masina:2020xhk, Baldes:2020nuv, Gondolo:2020uqv, Bernal:2020bjf}. Heavier PBHs, around stellar mass and higher, can instead contribute to the LIGO-Virgo gravitational wave (GW) events~\cite{Sasaki:2016jop, Bird:2016dcv, Clesse:2016vqa, Hutsi:2020sol, Hall:2020daa, Franciolini:2021tla, He:2023yvl} or provide seeds for cosmic structures~\cite{1983ApJ...275..405F, 1983ApJ...268....1C, Carr:2018rid, Liu:2022bvr, Hutsi:2022fzw}. 

The standard PBH formation scenarios are based on generation of peaks in the curvature power spectrum during cosmic inflation, which eventually leads to PBH formation as the highly overdense regions re-enter the horizon after inflation~\cite{Carr:1975qj}. As the resulting PBH abundance is exponentially sensitive to the amplitude of the perturbations, it seems difficult to avoid strong finetuning in such models. Other scenarios, such as PBH formation from collapsing false vacuum bubbles~\cite{Deng:2017uwc, Deng:2020mds, Kusenko:2020pcg, Maeso:2021xvl}, may in that regard look more attractive. In this paper, we will focus on PBH formation in first-order phase transitions. This was first suggested in~\cite{Hawking:1982ga,Kodama:1982sf} and recently studied further in~\cite{Lewicki:2019gmv, Kawana:2021tde, Liu:2021svg, Jung:2021mku, Hashino:2022tcs, Ashoorioon:2020hln, Huang:2022him, Kawana:2022lba, Kawana:2022olo}. 

First-order phase transitions proceed by the nucleation of bubbles of the broken phase in an initial background of the symmetric phase~\cite{Coleman:1977py,Callan:1977pt,Linde:1981zj}. These bubbles then expand until they collide and finish the conversion of the entire Universe. The growth of the bubbles transfers the energy of the false vacuum into the kinetic and gradient energy of the bubble walls. We focus on strong transitions where we can neglect the preexisting radiation. The key feature of such transitions is that the vacuum energy remains constant with the expansion of the Universe while the energy density of both bubble walls and radiation quickly decrease. By the statistical nature of bubble nucleation there is a chance of the existence of large regions filled with the false vacuum that are surrounded by regions in the true vacuum state. Such regions quickly become overdense because of the above-mentioned scaling with the expansion, and, if the generated overdensity is large enough, these regions may collapse into PBHs. 

We perform a detailed scan of the possible PBH formation scenarios involving such a strong transition. We go beyond the state of the art including firstly an additional term in the bubble nucleation rate which allows us to model very strong transitions in which the nucleation peaks and is cut off at later times. We also include the contribution to the energy of the false vacuum coming from the bubble walls which become very energetic in strong transitions. Finally, we include both large false vacuum regions that collapse immediately upon crossing the Hubble horizon and smaller regions that collapse as their compactness grows while they evolve already inside the horizon. We find that overproduction of PBHs constrains the bubble nucleation rate provided that the produced black holes are not light enough to evaporate before Big-Bang nucleosynthesis. We also show the impact of the nucleation history on the PBH mass function. The result is non-trivial as the small regions collapsing within the horizon form an additional peak of lighter PBHs which dominates in cases where the peak of bubble nucleation is reached. The separation between the peaks is, however, only by a factor of $\approx 2$, and the PBH mass function we find is always very narrow.

\section{Phase transition dynamics}

We adopt the thin-wall approximation to describe the growth of a single true vacuum bubble treating its wall as a boundary of negligible width between the domains of true and false vacuum with a certain surface energy density. Moreover, we consider the case where the interactions of the bubble wall with the surrounding particles can be neglected. This case is realized if the transition is strongly supercooled and the couplings of the particles with the scalar that drives the phase transition are small~\cite{Bodeker:2009qy,Bodeker:2017cim,Hoche:2020ysm,Gouttenoire:2021kjv,Sagunski:2023ynd,Lewicki:2022nba,Ellis:2020nnr}. With these approximations, the bubble is described by the action~\cite{Maeso:2021xvl}\footnote{We use the natural units $c = \hbar = 1$ throughout the paper.}
\begin{equation} \label{membrane_L}
    S = \int \td \eta \,a \left[ \frac{4\pi}{3} (a R)^3 \Delta V - 4\pi \sigma (a R)^2 \sqrt{1-(\partial_\eta R)^2} \right] = \int \td \eta \,\mathcal{L}(\eta,R,\partial_\eta R) \,,
\end{equation}
where $\eta$ denotes the conformal time, $a$ the scale factor, $R$ the bubble radius, $\sigma$ the surface tension of the bubble and $\Delta V$ the vacuum energy difference between false and true vacuum. The corresponding total energy of the bubble is given by
\begin{equation} \label{bubble_tot_energy}
    E = \frac1a \frac{\partial\mathcal{L}}{\partial (\partial_\eta R)} \partial_\eta R- \frac1a \mathcal{L} = -\frac{4\pi}{3}  (a R)^3 \Delta V + 4\pi  (a R)^2 \gamma \sigma \,,
\end{equation}
where the $1/a$ factors arise from the inverse metric and $\gamma = 1/\sqrt{1-(\partial_\eta R)^2}$ is the Lorentz factor of the bubble wall. We identify the second term in \eqref{bubble_tot_energy} with the bubble wall energy $E_w$, while the first term can be treated as the bubble volume energy. 

After the bubble nucleates, its growth is described by the equation of motion resulting from the above Lagrangian:
\begin{equation} \label{eom}
    \partial_\eta^2 R + 2\frac{1-(\partial_\eta R)^2}{R} \left(1 - \frac32 \frac{R}{R_H} \partial_\eta R \right) = \frac{a \Delta V}{\sigma} \left[1-(\partial_\eta R)^2\right]^{3/2}\,,
\end{equation}
where $R_H = (aH)^{-1} = a/\dot a$ denotes the comoving Hubble horizon radius. Given that we have neglected the friction terms, after a short accelerating period the bubble wall velocity approaches the speed of light, $\partial_\eta R \approx 1$, and the bubble radius increases linearly with the conformal time, 
\begin{equation} \label{radius}
    R(t_n,t) \approx \eta(t) - \eta(t_n) = \int_{t_n}^t \frac{\td t'}{a(t')} \,,
\end{equation}
where $t_n$ is the moment when the bubble nucleates. However, for the computation of the surface energy of the bubble, we need to accurately solve the $\gamma$ factor. The equation of motion can be rewritten in terms of $\gamma$ as
\begin{equation} \label{eq:gammaeom}
    \frac{\td \gamma}{\td R} + \frac{2\gamma}{R} \left(1 + \frac32 \frac{R}{R_H} \sqrt{1-\frac{1}{\gamma^2}} \right) = \frac{a \Delta V}{\sigma} \,.
\end{equation}
For $R < R_H$, using the initial condition $\gamma(R_n) = 1$, we get
\begin{equation}\label{surface_density}
	\gamma\sigma \approx \frac13 \Lambda \Delta V a R \,,\qquad  \Lambda \equiv \frac{(1+w)a + (1-w)a_n}{(3+w)a - (1+w)a_n} \,.
\end{equation}
with $w$ being the background equation of state parameter and $a_n \equiv a(\eta_n)$ the scale factor at the moment of bubble nucleation. We find that without the fitted $\Lambda$ factor, Eq.~\eqref{surface_density} underestimates $\gamma\sigma$, e.g. in radiation dominated background by $\sim 40\%$ when the radius of the bubble reaches the Hubble horizon radius. However, this is a good approximation only for a constant $w$. In the following, we solve numerically the equation of motion~\eqref{eom} as in the case of supercooled phase transitions $w$ changes during the period of bubble evolution.

The bubble nucleation is characterized by the nucleation rate per unit time and physical volume $\Gamma(t)$, which, in principle, is calculated using the Euclidean action for the critical bubble in a specific model~\cite{Coleman:1977py,Callan:1977pt,Linde:1981zj}. In order to perform a model independent analysis, we consider a generic bubble nucleation rate $\Gamma(t) = C e^{A(t)}$ and expand $A(t)$ to quadratic order around a time $t_0$, $A(t) = A_0 + \beta (t-t_0) - \zeta^2 (t-t_0)^2/2 + \mathcal{O}(t-t_0)^3$. By defining $\Gamma_0 \equiv C e^{A_0}$, can write the nucleation rate as
\begin{equation} \label{eq:gammaexp}
    \Gamma(t) = \Gamma_0 e^{\beta (t-t_0) - \frac12 \zeta^2 (t-t_0)^2} \,.
\end{equation} 
This approximation is relevant for strongly supercooled phase transitions and $\beta$ is the usual parameter describing timescale of the transition. The quadratic term in the exponent is introduced to mimic the effect of a temperature independent barrier in the potential that suppresses the bubble nucleation at late times. 

The bubble nucleation begins roughly when $\Gamma = H^4$. We assume that before this, the false vacuum energy, $\rho_v = \Delta V$, starts to dominate over the radiation energy density $\rho_r$. The Hubble rate $H$ then reaches a constant value
\begin{equation} \label{H_init}
    H_I^2 = \frac{\Delta V}{3 M_P^2} \,,
\end{equation}
where $M_P \equiv 8\pi/\sqrt{G}$ denotes the reduced Planck mass. After the bubble nucleation begins, the transition typically proceeds relatively quickly and therefore we choose the time $t_0$ around which the nucleation rate is expanded as the time when $\Gamma = H^4$. Moreover, without loss of generality, we choose $t_0=0$ so that $\Gamma_0 = H_I^4$. The rate~\eqref{eq:gammaexp} has a maximum at $t = \beta/\zeta^2 \equiv t_*$, and exchanging $\zeta^2$ for $t_*$ we can write the rate as
\begin{equation}\label{eq:Gamma}
    \Gamma(t) = H_I^4 e^{\beta t \left[1 - \frac{t}{2t_*}\right]} \,.
\end{equation}

The progression of the transition is characterized by the probability of finding a given point in the false vacuum~\cite{Guth:1982pn,Ellis:2018mja}
\begin{equation} \label{probability}
    P(t) = e^{-I(t)} \,, \qquad I(t) = \frac{4\pi}{3} \int_{-\infty}^t \td t_n \Gamma(t_n) a(t_n)^3 R(t_n;t)^3 \,.
\end{equation}
When the bubble nucleation begins, the vacuum energy density starts to decrease, $\rho_v = P(t) \Delta V$, as the bubbles convert the vacuum energy into energy of the bubble walls $\rho_w$. We can find how the wall energy scales with the expansion of the Universe using the continuity equation
\begin{equation}
    \frac{\td}{\td t} \frac{E}{a^3} + 3(1+w) H \frac{E_w}{a^3} = 0 \,,
\end{equation}
where $E$ is the bubble energy given in Eq.~\eqref{bubble_tot_energy}, $E_w$ the contribution of the bubble wall on $E$ and we have used the fact that $E-E_w$ scales as vacuum energy. Using the equation of motion~\eqref{eom} and neglecting the initial bubble size, we get $w = 1/3$.\footnote{This confirms earlier estimates based on lattice evolution of bubbles~\cite{Ellis:2019oqb}.} Therefore, the energy density of the bubble wall scales as radiation, $\rho_w \propto E_w/a^3 \propto a^{-4}$. The behavior of the scale factor is then determined by the Friedmann equations:
\begin{equation}\label{eq:Friedmann}
\begin{aligned}
     &H^2 = \frac{1}{3 M_P^2} \left(\rho_v + \rho_r + \rho_w \right) \,, \\
    &\frac{\td (\rho_r + \rho_w)}{\td t} +4 H(\rho_r + \rho_w) = - \frac{\td \rho_v}{\td t} \,.
\end{aligned}
\end{equation}
We label the initial values with I and, as in Eq.~\eqref{H_init}, assume the initial radiation density can be neglected, $\rho_r^I=0$. 

\begin{figure}
    \includegraphics[width=0.95\linewidth]{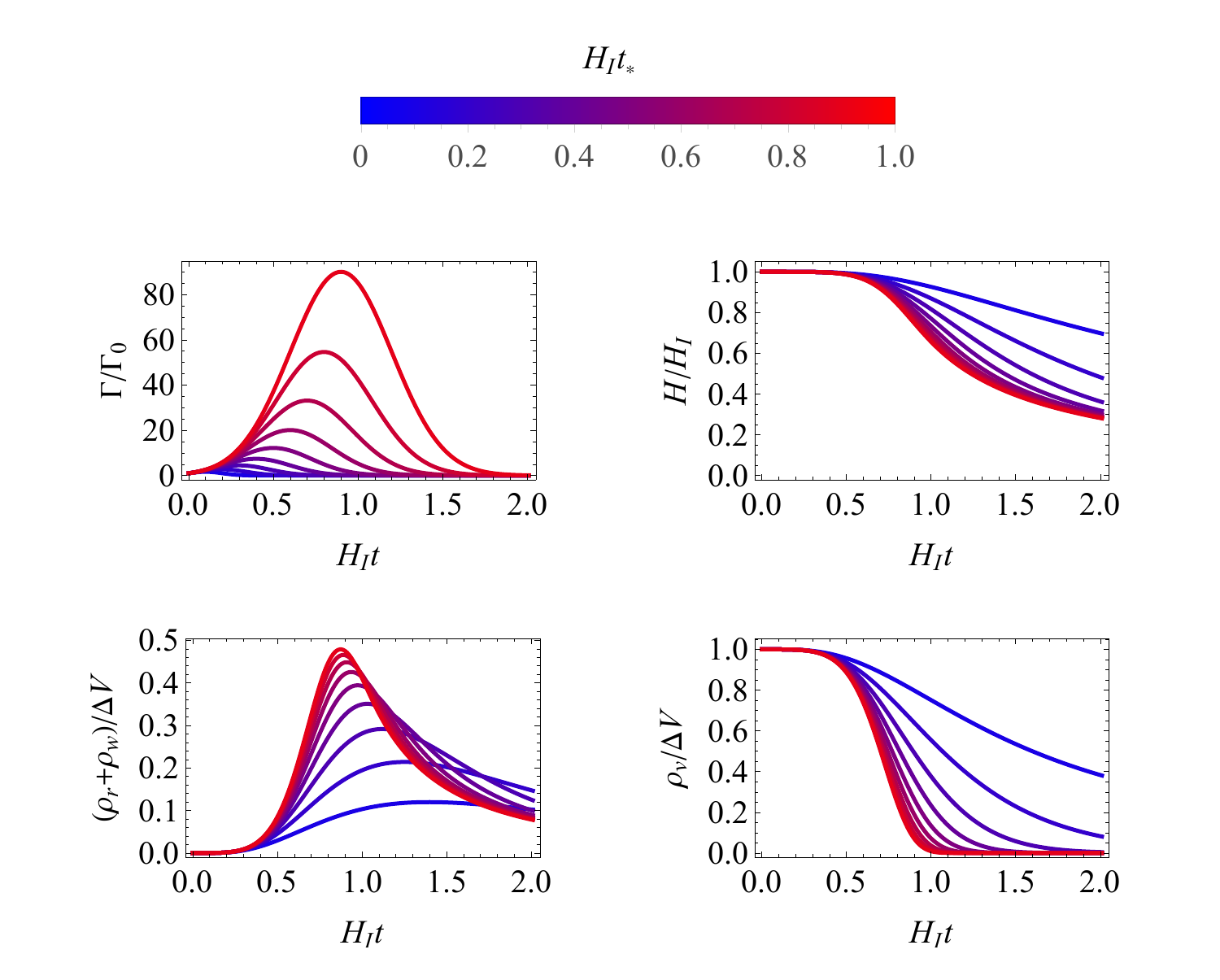}
	\caption{Quantities demonstrating dynamics of the transition including the vacuum decay rate, Hubble rate, energy density of walls and radiation and vacuum energy density. In this example we fixed $\beta/H_I = 10$, where $H_I$ is the initial value of Hubble parameter and varied $H_I t_*$  defining the peak of the nucleation rate.}
	\label{fig:Dynamics}
\end{figure}

In Fig.~\ref{fig:Dynamics} we show an example of the evolution of these quantities and the decay rate for various values of the time at which nucleation rate peaks normalised to the initial value of the Hubble parameter, $H_I t_*$ (see Eq.~\eqref{eq:Gamma}). For values close to unity the decay proceeds in a way similar to the usual case of pure exponential decay while smaller values cut the decay rate much faster and lead to slower conversion of the vacuum energy which might lead to problems with finishing the transition. The problem arises because patches of false vacuum undergo inflation. We have to require that the physical volume of false vacuum decreases for the transition to finish successfully~\cite{Turner:1992tz,Ellis:2018mja,Ellis:2019oqb}
\begin{equation}\label{criterion2}
	V_f^{-1}\frac{\td V_f}{\td t} = 3H - \frac{\td I}{\td t} < 0 \,,
\end{equation}
where $I(t)$ is given in Eq.~\eqref{probability}. In the following we chech that this condition is satisfied at late times.

\section{Trapped false vacuum domains}

Due to statistical fluctuations, there are regions of the Universe in which the transition into the true vacuum phase is postponed. At sufficiently late times, these regions may become so overdense that they experience gravitational collapse and become PBHs. In this section, following the reasoning described in Ref.~\cite{Kodama:1982sf}, we present a way to find distribution of such false vacuum domains (FVDs). 

\begin{figure}
\begin{center}
    \includegraphics[width=8cm]{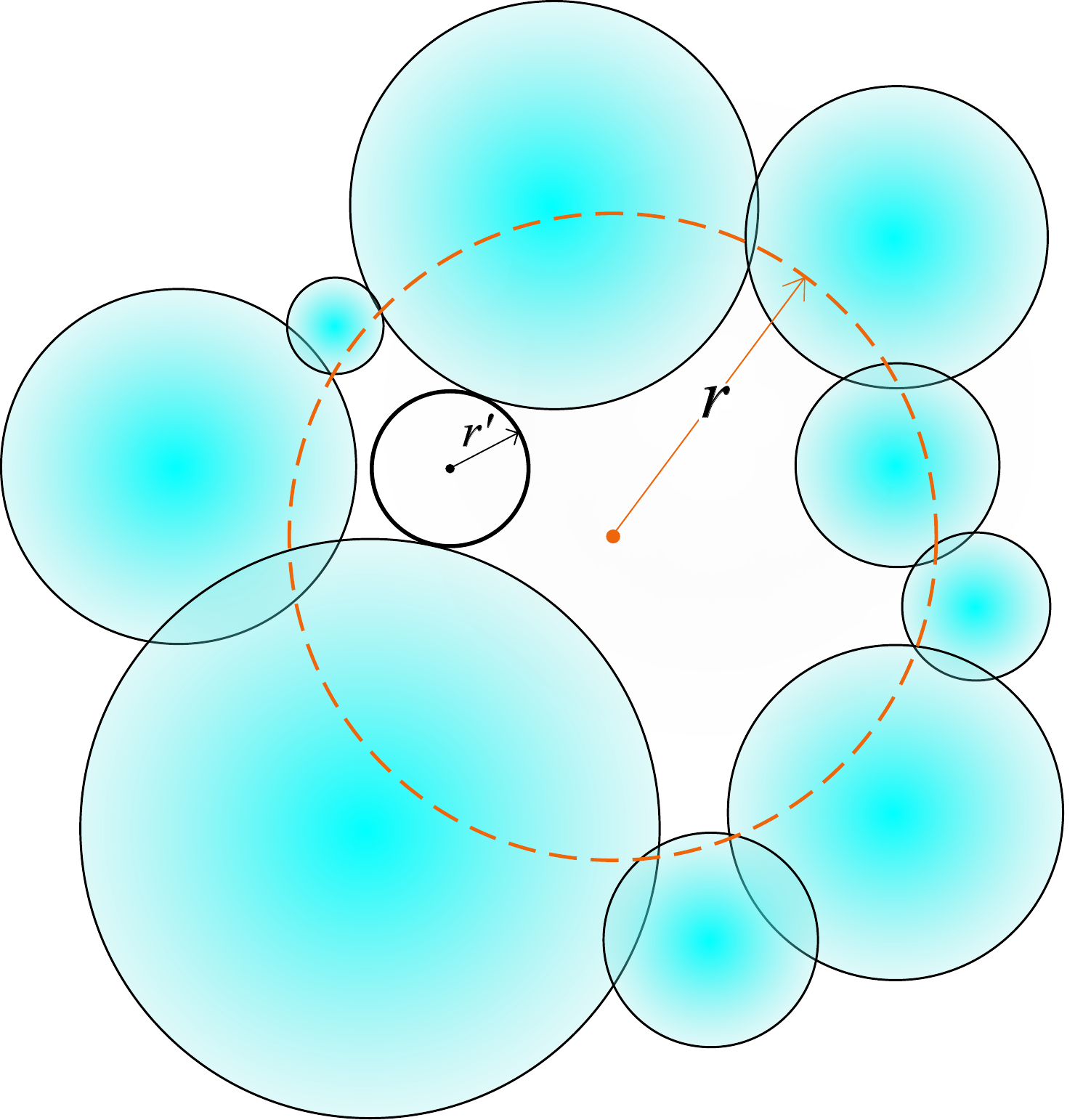}
    \caption{An schematic picture of a FVD. The quantities $r$ and $r'$ represent the radius of the FVD and the radius of a sphere that is completely filled with false vacuum.}
    \label{FVDs_fig}
\end{center}
\end{figure}

We denote the radius of the smallest sphere that encloses the FVD by $r$, as illustrated in Fig.~\ref{FVDs_fig}, and assume that the volume of the FVD is
\begin{equation}\label{FVD_vol}
	V(r, t) = \frac{4\pi}{3} a(t)^3 r^2(r - r_0(t)) \,,
\end{equation}
where $r_0(t)\geq 0$ parametrizes the non-sphericity of the FVD. This also indicates that there are no FVDs of radius $r<r_0(t)$. Furthermore, we introduce the function $\td n_{\rm FVD}(r,t)$, which gives the number density of FVDs whose radius is in the range $(r, r + \td r)$ at time $t$. 

We can relate $\td n_{\rm FVD}(r,t)$ with the probability $P(r',t)$ of a spherical domain of coordinate radius $r'$ to be filled completely with the false vacuum at time $t$,
\begin{equation}
    P(r',t) = \exp\left\{-\frac{4\pi}{3}\int_{-\infty}^t \td t_n \Gamma(t_n) a(t_n)^3 \left[r' +R(t_n,t)\right]^3\right\} \,.
\end{equation}
After the percolation time $t_p$, defined in the usual way as $P(t_p)=1/e$, any sphere of coordinate radius $r'$, that is filled completely with false vacuum, has to be contained inside some FVD. We must require that the center of the FVD is no further away from the center of the sphere than the coordinate distance $r - r'$. Also, $r'$ necessarily has to be smaller than the coordinate width $r-r_0(t)$ of the FVD, hence $r > r' + r_0(t)$. With these in mind, we come up with the following relation:
\begin{equation}\label{PF}
	P(r',t) = \frac{4\pi}{3} a(t)^3 \epsilon(t) \int_{r' + r_0(t)}^\infty \td r\, (r - r')^3 \frac{\td n_{\rm FVD}(r,t)}{\td r} \,.
\end{equation}
The factor $\epsilon(t)$ appearing here is introduced to take into account the non-sphericity of the FVDs. Eq.~\eqref{PF} has been solved using the Laplace transform in Ref.~\cite{Kodama:1982sf}. Here we give only the result:
\begin{equation}\label{eq:nFVD}
	\begin{split}
		\frac{\td n_{\rm FVD}(r)}{\td r} = & \frac{1}{\epsilon a^3} \frac{3}{4\pi}\frac{\Theta(r - r_0)}{r_0^3}\left[ -\frac{\td P(r - r_0)}{\td r} - \frac{3}{r_0}P(r - r_0)\right. \\
		& + \left. \frac{1}{r_0^2}\int_0^\infty \td\zeta P(r + \zeta - r_0)\left\{ \lambda e^{-\lambda\zeta/r_0} + \omega e^{-\omega\zeta/r_0} + \bar\omega e^{-\bar\omega\zeta/r_0}\right\} \right]\,,
	\end{split}
\end{equation}
where $\lambda$, $\omega$ and $\bar\omega$ are the roots of the polynomial $x^3 -3x^2 + 6x - 6$ and we have dropped off the time arguments for brevity. The function $r_0(t)$ is determined by requiring that $\td n_{\rm FVD}(r,t)/\td r$ is continuous at $r = r_0(t)$ and the function $\epsilon(t)$ by enforcing that the total volume fraction of false vacuum expressed with the use of $\td n_{\rm FVD}(r,t)/\td r$ coincides with $P(t)$:
\begin{equation}\label{FVD_vol2}
 	P(t) = \int_{r_0}^\infty \td r\, V(r, t) \frac{\td n_{\rm FVD}(r,t)}{\td r} \,,
\end{equation}
where the volume of the false vacuum inside the FVD of radius $r$ is given by Eq.~\eqref{FVD_vol}. 

\begin{figure}
    \centering
    \includegraphics[width=.462\textwidth]{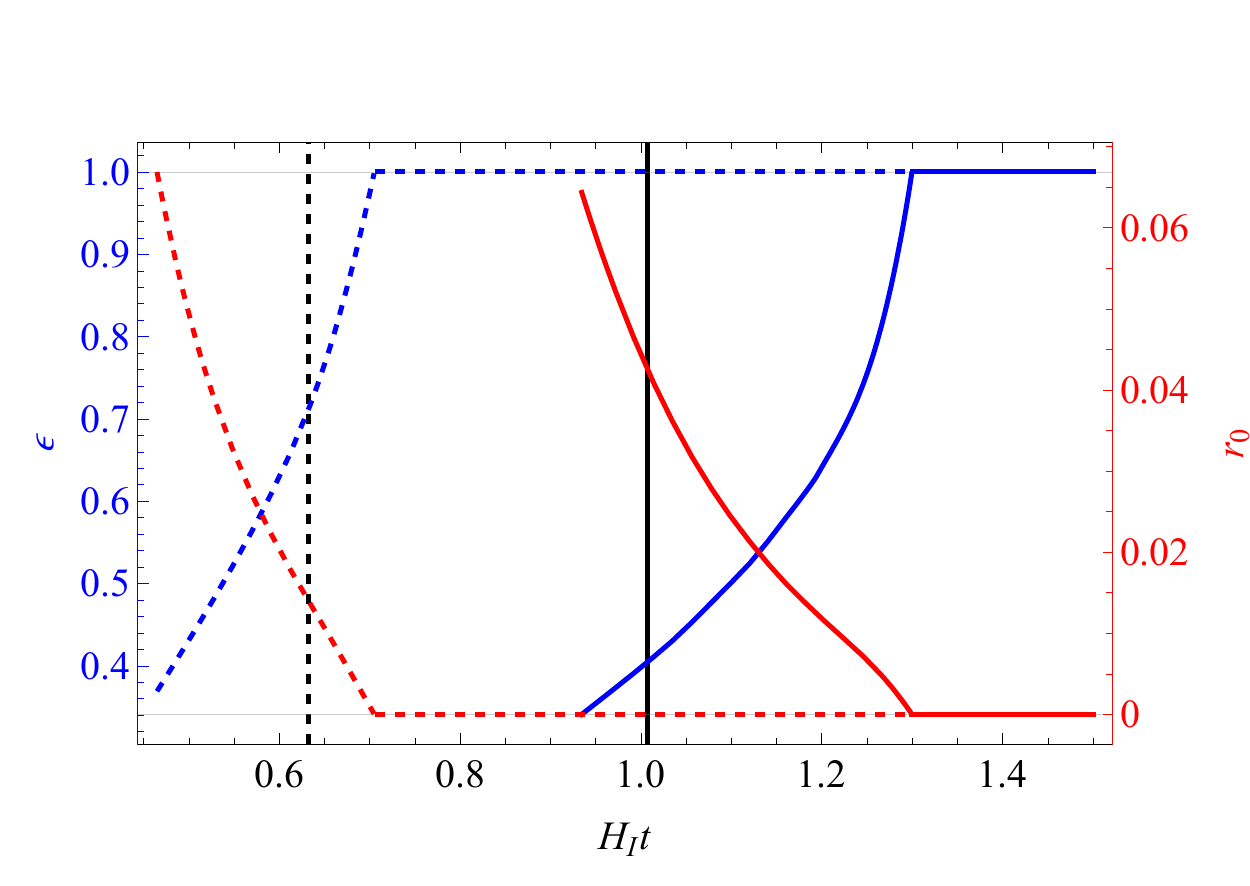} \hspace{1mm}
    \includegraphics[width=.512\textwidth]{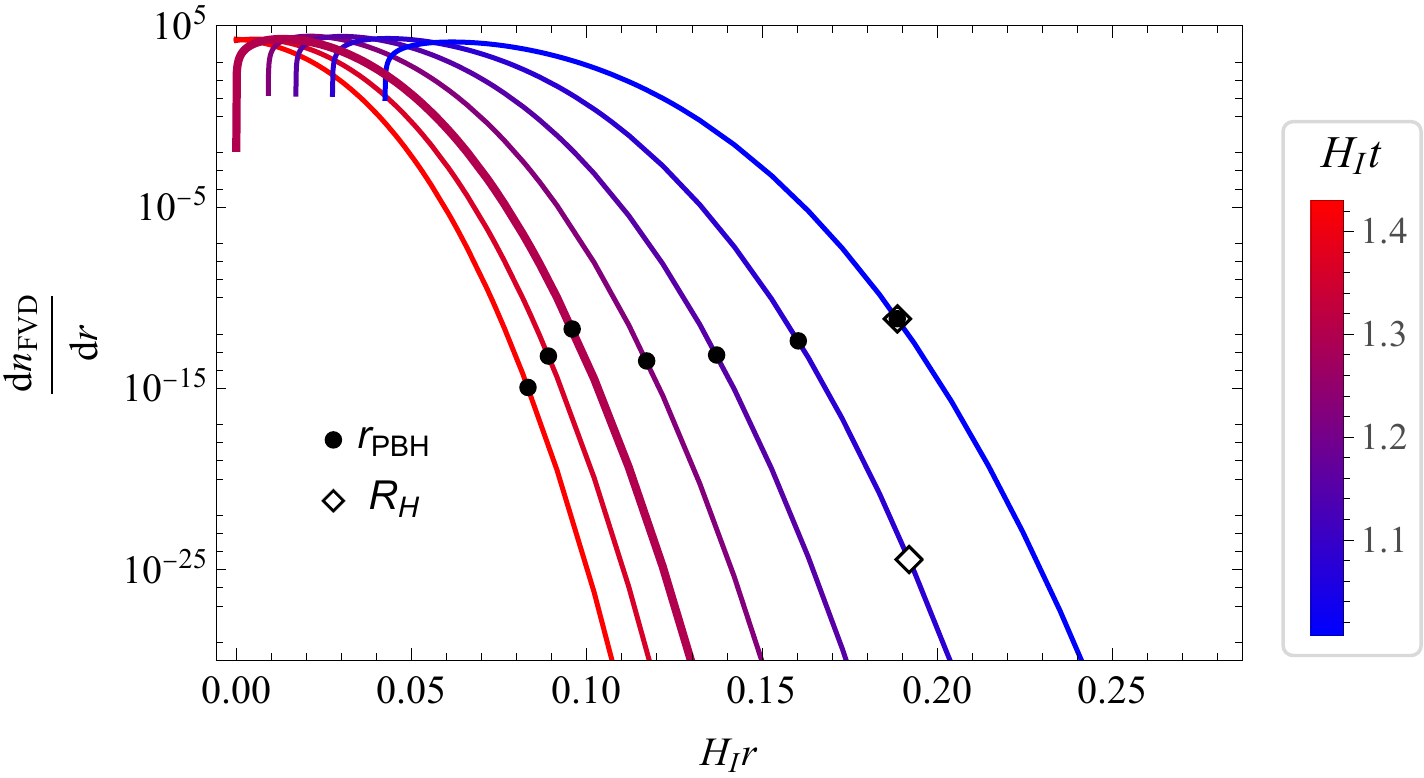}
	\caption{The left panel shows the non-sphericity parameters $r_0$ and $\epsilon$ as a function of time for two benchmark points: $\beta/H_I = 5$, $H_I t_* = 1.25$ (solid lines) and $\beta/H_I = 50$, $H_I t_* = 0.175$ (dashed lines). Black vertical lines denote the time $t_\times$ at which the hoop conjecture begins to be satisfied for regions smaller than the horizon radius $R_H$. The right panel shows the FVD distribution at different times after $t_\times$ (blue line). Thicker line represents the distribution at the time when sphericity of FVDs is reached. The black markers indicate the value of distribution for collapsing regions of radius $r_{\rm PBH}$ (circles) or $R_H$ (diamonds). The parameters in this benchmark case are $\beta/H_I = 5$, $H_I t_* = 1.25$.}
 \label{eps_fig}
\end{figure}

As shown in the left panel of Fig.~\ref{eps_fig}, $r_0(t)$ decreases and $\epsilon(t)$ increases with time. Eventually, $r_0(t)$ vanishes and $\epsilon(t)$ reaches $1$. This means that the FVDs are enclosed by so many true vacuum bubbles, that the false vacuum regions are approximately spherical. In this limit, the formula for the distribution $\td n_{\rm FVD}(r,t)/\td r$ takes a simplified form:
\begin{equation}
	\frac{\td n_{\rm FVD}(r, t)}{\td r} = a(t)^{-3} \frac{1}{8 \pi}\partial_r^{(4)}P(r, t) \,.
\end{equation}
We show the distribution $\td n_{\rm FVD}(r,t)$ at different times in the right panel of Fig.~\ref{eps_fig} in a benchmark case. We see that the large-$r$ tail of the distribution falls faster than $\propto e^{-H_I r}$. We will discuss this plot more in the end of the next section.

\section{Collapse into black holes} \label{sec:collapse}

We determine whether certain FVDs collapse to BHs by the hoop conjecture.\footnote{In an expanding background the hoop conjecture is slightly modified~\cite{Saini:2017tsz}, but these modifications have been shown to be insignificant~\cite{deJong:2021bbo}.} Since a uniform, constant density does not exert any net gravitational force, it plays no role in the collapse. We define the excess mass as the difference between total mass inside the sphere, $m(r,t)$, and the average mass of the region of the same size:
\begin{equation}\label{overdensity_mass}
	\delta m(r, t) \equiv m(r, t) - \frac{4\pi}{3}r^3a(t)^3\rho_B(t) \,,
\end{equation}
where $\rho_B(t)$ is the background energy density, which by the Friedmann equations~\eqref{eq:Friedmann} is given by
\begin{equation}\label{background}
	\rho_B(t) \equiv \rho_v(t) + \rho_r(t) + \rho_w(t) =  \Delta V \left[P(t) - \int_{-\infty}^t \!\td t'\, \frac{\td P}{\td t'} \left(\frac{a(t')}{a(t)}\right)^{\!4} \right] \,.
\end{equation}
The mass $m(r,t)$ consists of the following three contributions:
\begin{enumerate}
\item Mass of the false vacuum region occupying the fraction $\epsilon$ of FVD:
\begin{equation}\label{deltam1}
	m_1(r, t) = \frac{4\pi}{3}\epsilon(t) a^3(t)r^3(t) \Delta V \,.
\end{equation}
\item Mass of bubble walls:
\begin{equation}\label{deltam2}
	m_2(r, t) = 4\pi a^2(t)r^2(t) \sigma\bar\gamma(t) \,,
\end{equation}
where $\bar\gamma(t)$ denotes the average Lorentz factor of the true vacuum bubble walls.
\item Mass of true vacuum region:
\begin{equation}\label{deltam3}
    m_3(r, t) = \frac{4\pi}{3}(1-\epsilon(t))a^3(t)r^3(t)\rho_B(t) \,.
\end{equation}
\end{enumerate}
We approximate the contribution~\eqref{deltam2} from the bubble walls using the surface area of the smallest sphere enclosing the FVD. We thus do not fully include the real shape of the FVD which is composed of fragments of walls of true vacuum bubbles of different radii. We estimate $\bar\gamma(t)$ by solving numerically Eq.~\eqref{eom} with initial condition $\gamma(R_0)=1$ at the nucleation time of a bubble whose radius (evaluated using (\ref{radius})) at time $t$ equals the weighted average bubble radius\footnote{The weighting by $R^2$ arises from the assumption that the contribution from each bubble to the surface of the FVD scales with its surface area.}\footnote{This way of estimating $\bar\gamma(t)$ matches well with the more direct computation by $\bar\gamma(t) \propto \int^{t} \td n(t_n,t) R(t_n,t)^2 \gamma(t_n,t)$, but is numerically less expensive.}
\begin{equation}
 	\bar R(t) = \frac{\int_{-\infty}^{t} \td n(t_n,t) R(t_n,t)^3}{\int_{-\infty}^{t} \td n(t_n,t) R(t_n,t)^2} \,,
\end{equation}
where $\td n(t_n,t)$ denotes number density of true vacuum bubbles that nucleated within the time range $(t_n, t_n + \td t_n)$,
\begin{equation}\label{n(r,t)}
    \td n(t_n,t) = \frac{a^3(t_n)}{a^3(t)}P(t_n)\Gamma(t_n) \td t_n\,.
\end{equation}

We sum up all the contributions listed above, $m(r, t) = m_1(r, t)+m_2(r, t)+m_3(r, t)$, and arrive to the following formula for the excess mass of the FVD:
\begin{equation}\label{m_O(r,t)}
	\delta m(r, t) = \frac{4\pi}{3}a^3(t)r^3\epsilon(t)\left(\Delta V - \rho_B(t)\right) + 4 \pi r^2 a^2(t) \sigma \bar\gamma(t) \,.
\end{equation}
The hoop conjecture then implies that if the radius $r$ 
of the FVD smaller than the Schwarzschild radius of the excess mass $\delta m$, the FVD will collapse to a BH:
\begin{equation}\label{Schwarzschild_cond}
    r < r_s(\delta m) = \frac{\delta m(r,t)}{4\pi a(t) M_P^2} \,.
\end{equation}
Using Eq.~\eqref{m_O(r,t)}, the condition~\eqref{Schwarzschild_cond} can be recast as a lower bound on $r$:
\begin{equation}\label{r_PBH}
	r > r_{\rm PBH}(t) \equiv 2 R_H \sqrt{\frac{\rho_B(t)}{\Delta V}} \left[ \sqrt{3} \frac{\sigma \bar\gamma(t)}{M_P \sqrt{\Delta V}} + \sqrt{4\epsilon(t)\left(1-\frac{\rho_B(t)}{\Delta V}\right) + 3 \frac{\sigma^2 \bar\gamma(t)^2}{M_P^2 \Delta V}} \right]^{-1} \,,
\end{equation}
where $R_H(t) = 1/(a(t) H(t))$ is the comoving Hubble radius. We note that $\epsilon(t)$, and the fractions $\rho_B(t)/\Delta V$ and $\sigma \bar\gamma(t)/\sqrt{\Delta V}$ are independent of $\Delta V$,\footnote{Using $x\equiv H_I R$, the equation of motion~\eqref{eq:gammaeom} can be written for $\gamma \gg 1$ as
\begin{equation*}
    \left[\frac{\td}{\td x} + \frac{2}{x} \left(1 + \frac{3}{2} \sqrt{\frac{\rho_B}{\Delta V}} a x \right)\right] \frac{\sigma\gamma}{M_P \sqrt{\Delta V}} = 3 a\,.
\end{equation*}
} so also the fraction $r_{\rm PBH}(t)/R_H(t)$ is independent of $\Delta V$.

\begin{figure}
    \centering
    \includegraphics[width=.44\textwidth]{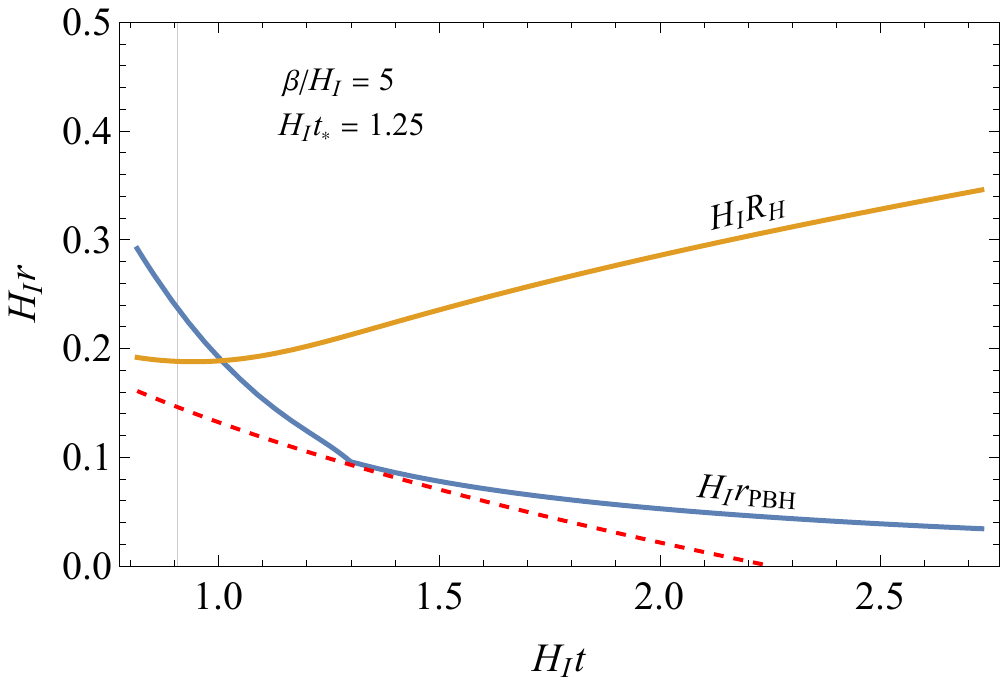} \hspace{2mm}
    \includegraphics[width=.44\textwidth]{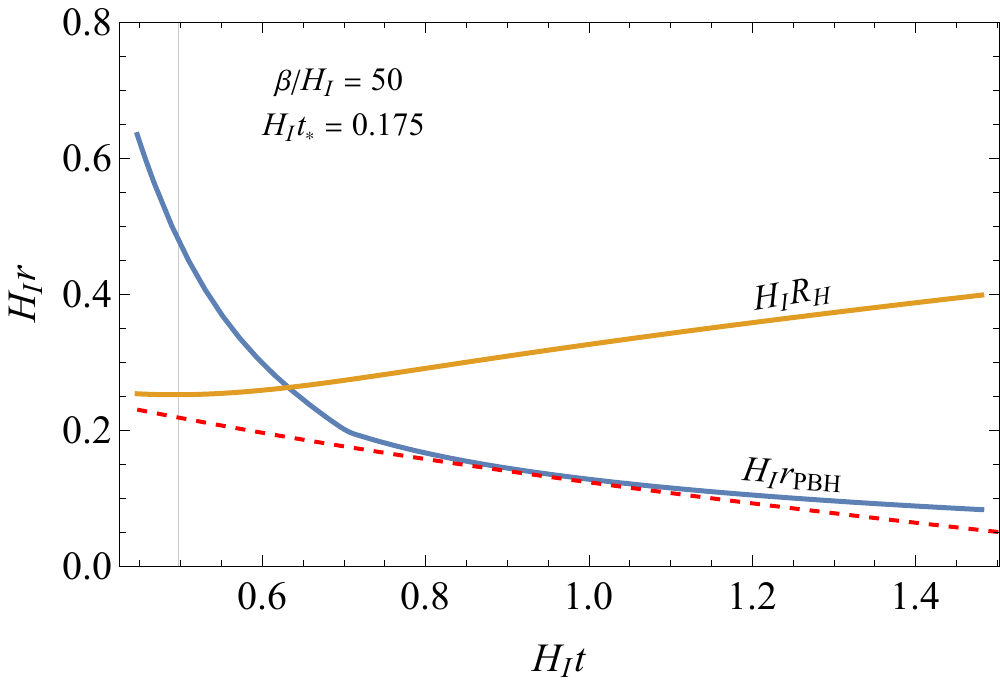}
	\caption{Evolution of lower bound on radius of collapsing regions (blue curve) and comoving Hubble radius (orange curve) during the transition. The dashed red curves show the radius of a false vacuum domain that at one moment reaches $r_{\rm PBH}$ but never becomes larger than it. The vertical gray line indicates the percolation moment, $t = t_p$.}
 \label{rpbh_fig}
\end{figure}

As shown for two representative benchmark cases in Fig.~\ref{rpbh_fig}, $r_{\rm PBH}$ decreases with time and becomes smaller than the horizon radius $R_H$ soon after percolation time $t=t_p$.\footnote{The kink in the $r_{\rm PBH}$ curve seen in both cases after the Hubble crossing corresponds to the moment when the false vacuum domains become spherical, $\epsilon(t) = 1$ (cf. Fig.~\ref{eps_fig}).} We denote the Hubble crossing moment by $t_\times$, ie. $r_{\rm PBH}(t_\times)=R_H(t_\times)$. After that moment, all false vacuum regions entering the Hubble horizon will collapse to PBHs. In addition, any subhorizon regions that reach $r>r_{\rm PBH}$ can collapse to PBHs. We remark, that as shown in Ref.~\cite{Lewicki:2023mik}, interactions of the walls with the surrounding particles whose mass increases in the phase transition can prevent the PBH formation from sub-horizon false vacuum bubbles as the increasing density of the particles can eventually stop the wall. This does not, however, affect the regions that are sufficiently overdense already when entering the Hubble horizon.

The red dashed curves in Fig.~\ref{rpbh_fig} show the radius of a false vacuum domain that at one moment reaches $r_{\rm PBH}$ but never becomes larger than it, computed assuming that it is a sphere bounded whose radius increases according to Eq.~\eqref{radius} and neglecting possible bubble nucleation inside it. In order to limit double counting, we neglect the part of blue line for which the slope is less steep than the slope of red line, since these cases are expected to correspond regions that have already collapsed.

We show the distribution of the FVDs, $\td n_{\rm FVD}/\td r$, at times after $t_\times$ in a benchmark case in the right panel of Fig.~\ref{eps_fig}. The circles along each of the curves indicate $r_{\rm PBH}$ and the diamonds $R_H$. At $t=t_\times$ (blue curve) these coincide. We see that the PBHs form from very rare FVDs, far in the large-$r$ tail of the distribution.

\section{Results}

We perform numerical computations scanning the parameter space with $1 < \beta/H_I < 100$ and $0.03 <  H_I t_* < 10$. At each point we first check whether the percolation condition~\eqref{criterion2} is satisfied. If so, we proceed to the computation of PBH formation. The mass function of the PBHs consists of two parts, horizon size regions that collapse at horizon entry and sub-horizon regions that cross the threshold:
\begin{equation}
\begin{aligned}
    \frac{\td n_{\rm PBH}(t)}{\td m} = \int_{t_\times}^t \td t' \,\left[\frac{a(t')}{a(t)}\right]^3 \bigg[&\frac{\td R_H}{\td t'} \frac{\td n_{\rm FVD}(R_H,t')}{\td r} \delta(m-m(R_H,t')) \\
    &+ \frac{\td r_{\rm PBH}}{\td t'} \frac{\td n_{\rm FVD}(r_{\rm PBH},t')}{\td r} \delta(m-m(r_{\rm PBH},t')) \bigg] \,.
\end{aligned}
\end{equation}
The total PBH energy density $\rho_{\rm PBH}$ and their mass function $\psi(m)$, normalized so that $\int \td m \, \psi(m) = 1$, are then given by
\begin{equation}
    \rho_{\rm PBH} = \int \td m \, m \frac{\td n_{\rm PBH}}{\td m} \,, \qquad
    \psi(m) = \frac{m}{\rho_{\rm PBH}} \frac{\td n_{\rm PBH}}{\td m} \,,
\end{equation}
and the present day fraction of DM in PBHs, assuming they don't evaporate by today, is given by $f_{\rm PBH} = \rho_{\rm PBH}/\rho_{\rm DM}$, where $\rho_{\rm DM}$ denotes the DM abundance. We remark that $f_{\rm PBH}$ depends on $\Delta V$ only through the expansion of the Universe, that leads to the scaling $f_{\rm PBH} \propto \Delta V^{1/4}$ as the temperature after the transition is $T_{\rm reh} \propto \Delta V^{1/4}$.

\begin{figure}
\begin{center}
\includegraphics[width=0.7\textwidth]{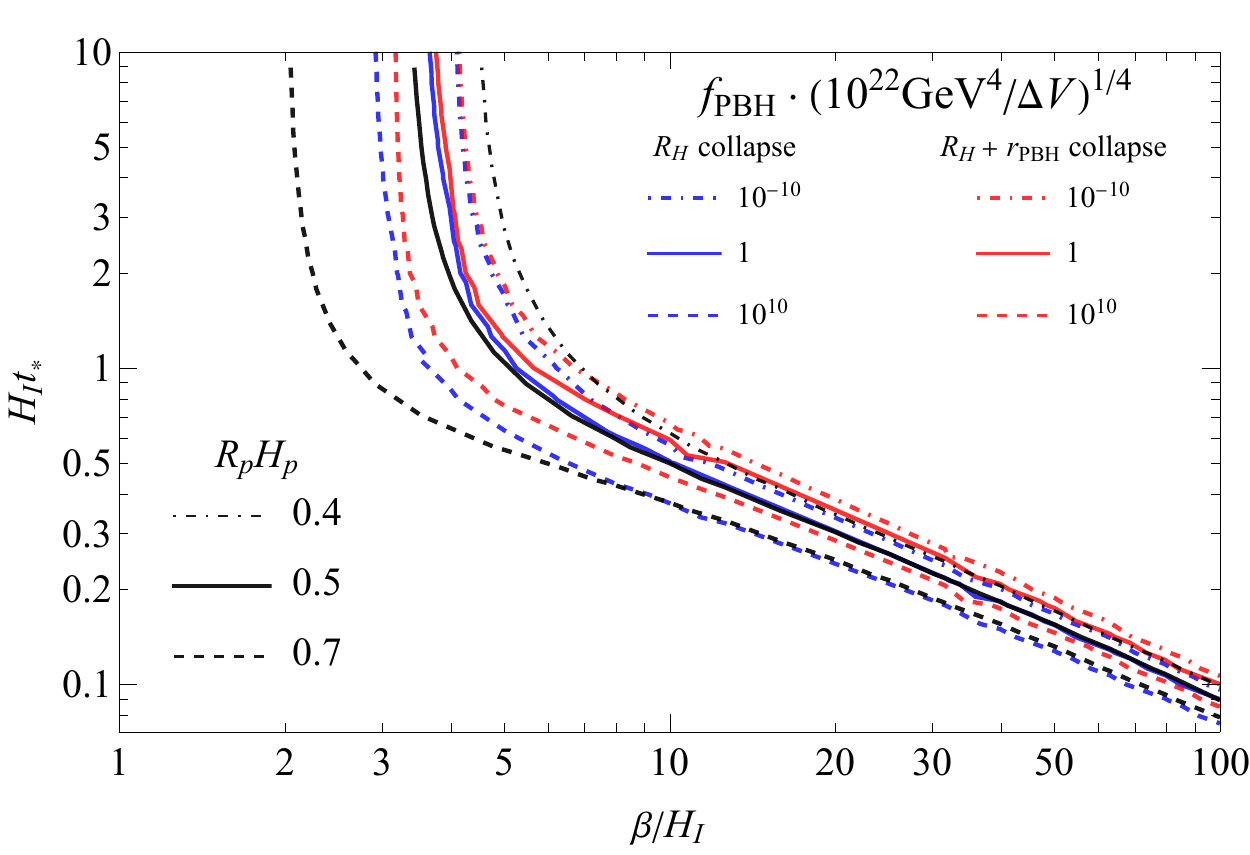}
\caption{The PBH abundance today normalised to that of DM, $f_{\rm PBH}$, as a function of the bubble nucleation parameters $\beta$ and $t_*$. The blue lines include only large regions collapsing immediately on horizon crossing while the red ones also include the population of smaller regions collapsing as their compactness grows. The black lines show the average bubble radius normalised to the horizon radius, $R_p H_p$.}
\label{fig:fpbh}
\end{center}
\end{figure}

The resulting $f_{\rm PBH}$ is shown as a function of $\beta/H_I$ and $H_I t_*$ in Fig.~\ref{fig:fpbh}. By comparing the colored curves with the black ones, we see that the abundance does not show a strong dependence of $\beta$ and $t_*$ separately, but is mainly determined by $R_p H_p$ where $H_p = H(t_p)$ is the Hubble rate at the percolation time and $R_p$ is the mean bubble radius~\cite{Turner:1992tz}
\begin{equation}
    R_p = \left[\int_{-\infty}^{t_p} \! {\rm d} t \, \frac{a(t)^3}{a(t_p)^3} \Gamma(t) P(t) \right]^{-1/3} .
\end{equation} 
A significant abundance of PBHs can be formed if the mean bubble radius is large, $R_p H_p \simeq 0.5$. We show the result both with and without the collapse of sub-horizon regions, and the real result lies somewhere between these approximation as the inclusion of the subhorizon regions may include double counting. We find, however, that the PBH abundance changes rapidly with $R_p H_p$ and therefore a small change in $R_p H_p$ can compensate the uncertainty in $f_{\rm PBH}$ arising from this double counting.

The value $R_p H_p \simeq 0.5$ corresponds to $\beta/H=3.8$ for approximately exponential bubble nucleation rate. Ref.~\cite{Liu:2021svg} found the threshold to be $\beta/H\approx9$ for the same conditions corresponding to $\alpha \gg 1$. It was, however, shown in~\cite{Kawana:2022olo} that they overestimated the survival probability of the false vacuum patches. The latter reference found the threshold to be $\beta/H=2.5$ for $\alpha \sim 1$. We find higher PBH formation efficiency because we consider $\alpha\gg 1$ and also collapse of subhorizon regions. In~\cite{Gouttenoire:2023naa}, that appeared after our paper, the threshold was found to be $\beta/H \approx 7$. The main difference compared to our work is that they consider the collapse of patches that can include true vacuum bubbles.

\begin{figure}
\begin{center}	\includegraphics[width=0.6\textwidth]{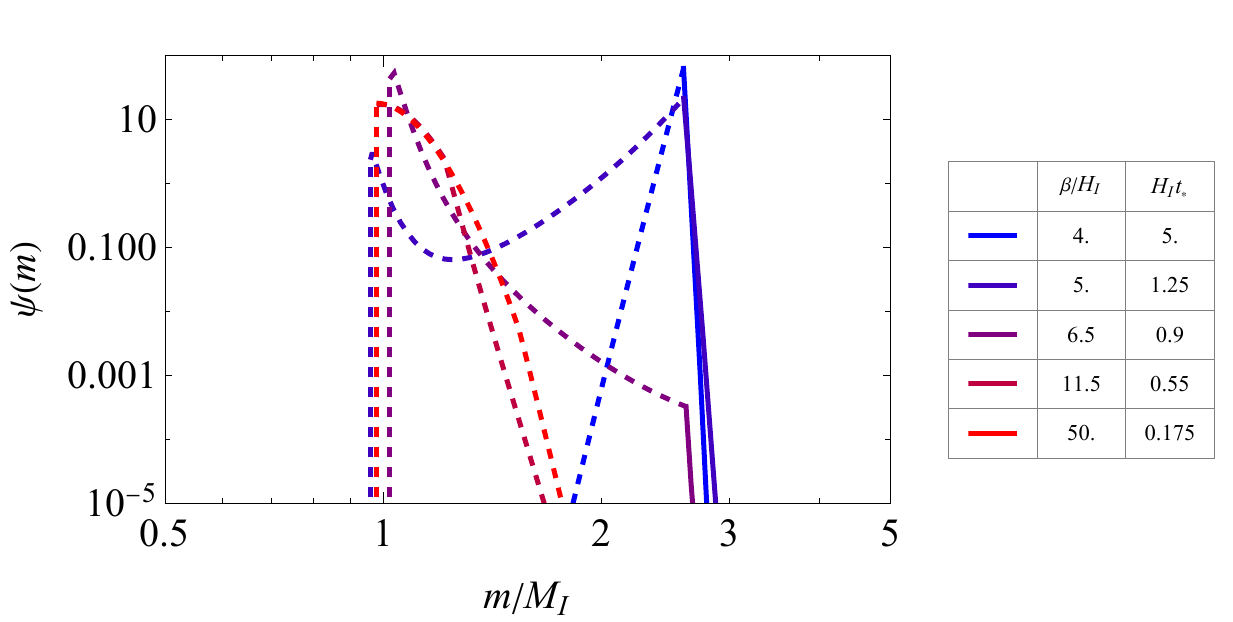}
\caption{The normalised PBH mass function produced for a set of transitions with different values of the bubble nucleation rate parameters $\beta$ and $t_*$. The examples here are chosen to lay along the $f_{\rm PBH}=1$ line for $\Delta V = 10^{22}\,{\rm GeV}^4$.}
\label{fig:spectra}
\end{center}
\end{figure}

The PBH mass function is shown in Fig.~\ref{fig:spectra}. The mass scale of the produced PBHs is given by the horizon mass:
\begin{equation}
    M_I = \frac{4\pi}{3} H_I^{-3} \Delta V = 4\sqrt{3}\pi M_P^3 \Delta V^{-1/2} \approx 0.09 M_\odot (\Delta V/{\rm GeV}^4)^{-1/2} \,.
\end{equation}
We see that in particular the low mass part of the mass function, that arises from the collapse of subhorizon regions, changes with $\beta$ and $t_*$ and can include two peaks. The PBH mass function is, however, in every case very narrow, ranging at most from $m \approx M_I$ to $m \approx 3 M_I$. In the case with $t_*\gg 0$, where $\Gamma \approx H_I^4 e^{\beta t}$, the mass function is approximately a delta function at $m \approx 2.6 M_I$.

From Fig.~\ref{fig:fpbh} we see that in part of the parameter space $f_{\rm PBH} > 1$. It is evident that in case of supercooled transitions, overproduction of PBHs becomes a strong condition which has to be taken into account. This bound can be avoided if the PBHs are sufficiently light and evaporate much before the Big-Bang nucleosynthesis (BBN). By requiring that the lifetime of a non-spinning BH of mass $m$~\cite{Hawking:1974rv,Hawking:1975vcx} $\tau \approx 0.4\,{\rm s} \,(m/10^9{\rm g})^3$, is less than a second, we find that the bound from PBH overproduction can be avoided for $\Delta V^{1/4} \gtrsim 10^{12}$\,GeV. However, the evaporation products and possible remnants can play the role of DM. On the other hand, in the asteroid mass range, $10^{-16} \lesssim m/M_\odot \lesssim 10^{-12}$, PBHs can constitute all DM~\cite{Carr:2020gox}. PBHs in that mass window can be generated in a strongly supercooled phase transition if $10^5\,{\rm GeV} \lesssim \Delta V^{1/4} \lesssim 10^8\,{\rm GeV}$.

\section{Conclusions}

We have estimated the population of PBHs produced in a strongly supercooled first-order phase transition. We have included both the contribution coming from large false vacuum regions which collapse upon crossing the Hubble horizon and the collapse of smaller subhorizon regions. We have modeled the transition with the decay rate that involves the standard exponential decay rate described by the usual $\beta/H_I$ term as well as the second order correction that produces a peak in the decay rate at some later time $H_I t_*$. This has allowed us to probe the parameter space of models realizing a very strong transition where purely exponential decay might not give an accurate description.

We have shown that the abundance of PBHs produced in the transition can roughly be approximated through the average bubble radius normalised to the horizon size. The abundance is very sensitive to $R_p H_p$, with $R_p H_p \approx 0.5$ resulting in a sizeable PBH abundance. In particular, for approximately exponential bubble nucleation rate the observed DM abundance in PBHs is produced for $\beta/H_I \approx 3.8$. 

The mass scale of the produced PBHs is given by the horizon mass at the beginning of the transition which is determined for strongly supercooled transitions by the potential energy difference $\Delta V$. We have shown that the overproduction of PBHs constraints the parameter space if $\Delta V < (10^{12} {\rm GeV})^4$, as in this case the PBHs don't evaporate before the Big-Bang nucleosynthesis and that the observed DM abundance can be formed in asteroid mass PBHs if $10^5\,{\rm GeV} \lesssim \Delta V^{1/4} \lesssim 10^8\,{\rm GeV}$.

The shape of the bubble nucleation rate affects the final PBH mass function. Large regions collapsing immediately upon horizon crossing give rise to the steep high mass slope of the final spectrum while the shape at lower masses is dictated by the smaller false vacuum remnants crossing the collapse threshold as they shrink. The latter contribution becomes particularly relevant if the bubble nucleation rate has a maximum soon after reaching the Hubble rate. For approximately exponential bubble nucleation rate we have found a very narrow PBH mass function peaked at mass $0.23 (\Delta V/{\rm GeV})$.

\section*{Acknowledgements}
This work was supported by the Polish National Agency for Academic Exchange within Polish Returns Programme under agreement PPN/PPO/2020/1/00013/U/00001 and the Polish National Science Center grant 2018/31/D/ST2/02048. The work of V.V. was supported by the European Union's Horizon Europe research and innovation program under the Marie Sk\l{}odowska-Curie grant agreement No. 101065736, by the European Regional Development Fund through the CoE program grant TK133 and by the Estonian Research Council grant PRG803.

\bibliographystyle{JHEP}
\bibliography{Bib}

\providecommand{\href}[2]{#2}\begingroup\raggedright\begin{thebibliography}{10}

\bibitem{Carr:2020gox}
B.~Carr, K.~Kohri, Y.~Sendouda and J.~Yokoyama, \emph{{Constraints on
  primordial black holes}},
  \href{https://doi.org/10.1088/1361-6633/ac1e31}{\emph{Rept. Prog. Phys.}
  {\bfseries 84} (2021) 116902}
  [\href{https://arxiv.org/abs/2002.12778}{{\ttfamily 2002.12778}}].

\bibitem{Hawking:1974rv}
S.W.~Hawking, \emph{{Black hole explosions}},
  \href{https://doi.org/10.1038/248030a0}{\emph{Nature} {\bfseries 248} (1974)
  30}.

\bibitem{Hawking:1975vcx}
S.W.~Hawking, \emph{{Particle Creation by Black Holes}},
  \href{https://doi.org/10.1007/BF02345020}{\emph{Commun. Math. Phys.}
  {\bfseries 43} (1975) 199}.

\bibitem{Fujita:2014hha}
T.~Fujita, M.~Kawasaki, K.~Harigaya and R.~Matsuda, \emph{{Baryon asymmetry,
  dark matter, and density perturbation from primordial black holes}},
  \href{https://doi.org/10.1103/PhysRevD.89.103501}{\emph{Phys. Rev. D}
  {\bfseries 89} (2014) 103501}
  [\href{https://arxiv.org/abs/1401.1909}{{\ttfamily 1401.1909}}].

\bibitem{Allahverdi:2017sks}
R.~Allahverdi, J.~Dent and J.~Osinski, \emph{{Nonthermal production of dark
  matter from primordial black holes}},
  \href{https://doi.org/10.1103/PhysRevD.97.055013}{\emph{Phys. Rev. D}
  {\bfseries 97} (2018) 055013}
  [\href{https://arxiv.org/abs/1711.10511}{{\ttfamily 1711.10511}}].

\bibitem{Lennon:2017tqq}
O.~Lennon, J.~March-Russell, R.~Petrossian-Byrne and H.~Tillim, \emph{{Black
  Hole Genesis of Dark Matter}},
  \href{https://doi.org/10.1088/1475-7516/2018/04/009}{\emph{JCAP} {\bfseries
  04} (2018) 009} [\href{https://arxiv.org/abs/1712.07664}{{\ttfamily
  1712.07664}}].

\bibitem{Hooper:2019gtx}
D.~Hooper, G.~Krnjaic and S.D.~McDermott, \emph{{Dark Radiation and Superheavy
  Dark Matter from Black Hole Domination}},
  \href{https://doi.org/10.1007/JHEP08(2019)001}{\emph{JHEP} {\bfseries 08}
  (2019) 001} [\href{https://arxiv.org/abs/1905.01301}{{\ttfamily
  1905.01301}}].

\bibitem{Masina:2020xhk}
I.~Masina, \emph{{Dark matter and dark radiation from evaporating primordial
  black holes}},
  \href{https://doi.org/10.1140/epjp/s13360-020-00564-9}{\emph{Eur. Phys. J.
  Plus} {\bfseries 135} (2020) 552}
  [\href{https://arxiv.org/abs/2004.04740}{{\ttfamily 2004.04740}}].

\bibitem{Baldes:2020nuv}
I.~Baldes, Q.~Decant, D.C.~Hooper and L.~Lopez-Honorez, \emph{{Non-Cold Dark
  Matter from Primordial Black Hole Evaporation}},
  \href{https://doi.org/10.1088/1475-7516/2020/08/045}{\emph{JCAP} {\bfseries
  08} (2020) 045} [\href{https://arxiv.org/abs/2004.14773}{{\ttfamily
  2004.14773}}].

\bibitem{Gondolo:2020uqv}
P.~Gondolo, P.~Sandick and B.~Shams Es~Haghi, \emph{{Effects of primordial
  black holes on dark matter models}},
  \href{https://doi.org/10.1103/PhysRevD.102.095018}{\emph{Phys. Rev. D}
  {\bfseries 102} (2020) 095018}
  [\href{https://arxiv.org/abs/2009.02424}{{\ttfamily 2009.02424}}].

\bibitem{Bernal:2020bjf}
N.~Bernal and O.~Zapata, \emph{{Dark Matter in the Time of Primordial Black
  Holes}}, \href{https://doi.org/10.1088/1475-7516/2021/03/015}{\emph{JCAP}
  {\bfseries 03} (2021) 015}
  [\href{https://arxiv.org/abs/2011.12306}{{\ttfamily 2011.12306}}].

\bibitem{Sasaki:2016jop}
M.~Sasaki, T.~Suyama, T.~Tanaka and S.~Yokoyama, \emph{{Primordial Black Hole
  Scenario for the Gravitational-Wave Event GW150914}},
  \href{https://doi.org/10.1103/PhysRevLett.117.061101}{\emph{Phys. Rev. Lett.}
  {\bfseries 117} (2016) 061101}
  [\href{https://arxiv.org/abs/1603.08338}{{\ttfamily 1603.08338}}].

\bibitem{Bird:2016dcv}
S.~Bird, I.~Cholis, J.B.~Mu\~noz, Y.~Ali-Ha\"\i{}moud, M.~Kamionkowski,
  E.D.~Kovetz et~al., \emph{{Did LIGO detect dark matter?}},
  \href{https://doi.org/10.1103/PhysRevLett.116.201301}{\emph{Phys. Rev. Lett.}
  {\bfseries 116} (2016) 201301}
  [\href{https://arxiv.org/abs/1603.00464}{{\ttfamily 1603.00464}}].

\bibitem{Clesse:2016vqa}
S.~Clesse and J.~Garc\'\i{}a-Bellido, \emph{{The clustering of massive
  Primordial Black Holes as Dark Matter: measuring their mass distribution with
  Advanced LIGO}},
  \href{https://doi.org/10.1016/j.dark.2016.10.002}{\emph{Phys. Dark Univ.}
  {\bfseries 15} (2017) 142}
  [\href{https://arxiv.org/abs/1603.05234}{{\ttfamily 1603.05234}}].

\bibitem{Hutsi:2020sol}
G.~H\"utsi, M.~Raidal, V.~Vaskonen and H.~Veerm\"ae, \emph{{Two populations of
  LIGO-Virgo black holes}},
  \href{https://doi.org/10.1088/1475-7516/2021/03/068}{\emph{JCAP} {\bfseries
  03} (2021) 068} [\href{https://arxiv.org/abs/2012.02786}{{\ttfamily
  2012.02786}}].

\bibitem{Hall:2020daa}
A.~Hall, A.D.~Gow and C.T.~Byrnes, \emph{{Bayesian analysis of LIGO-Virgo
  mergers: Primordial vs. astrophysical black hole populations}},
  \href{https://doi.org/10.1103/PhysRevD.102.123524}{\emph{Phys. Rev. D}
  {\bfseries 102} (2020) 123524}
  [\href{https://arxiv.org/abs/2008.13704}{{\ttfamily 2008.13704}}].

\bibitem{Franciolini:2021tla}
G.~Franciolini, V.~Baibhav, V.~De~Luca, K.K.Y.~Ng, K.W.K.~Wong, E.~Berti
  et~al., \emph{{Searching for a subpopulation of primordial black holes in
  LIGO-Virgo gravitational-wave data}},
  \href{https://doi.org/10.1103/PhysRevD.105.083526}{\emph{Phys. Rev. D}
  {\bfseries 105} (2022) 083526}
  [\href{https://arxiv.org/abs/2105.03349}{{\ttfamily 2105.03349}}].

\bibitem{He:2023yvl}
J.~He, H.~Deng, Y.-S.~Piao and J.~Zhang, \emph{{Implications of GWTC-3 on
  primordial black holes from vacuum bubbles}},
  \href{https://arxiv.org/abs/2303.16810}{{\ttfamily 2303.16810}}.

\bibitem{1983ApJ...275..405F}
K.~{Freese}, R.~{Price} and D.N.~{Schramm}, \emph{{Formation of population III
  stars and galaxies with primordial planetary-mass black holes}},
  \href{https://doi.org/10.1086/161542}{\emph{Astrophysical Journal} {\bfseries
  275} (1983) 405}.

\bibitem{1983ApJ...268....1C}
B.J.~{Carr} and J.~{Silk}, \emph{{Can graininess in the early universe make
  galaxies?}}, \href{https://doi.org/10.1086/160924}{\emph{Astrophysical
  Journal} {\bfseries 268} (1983) 1}.

\bibitem{Carr:2018rid}
B.~Carr and J.~Silk, \emph{{Primordial Black Holes as Generators of Cosmic
  Structures}}, \href{https://doi.org/10.1093/mnras/sty1204}{\emph{Mon. Not.
  Roy. Astron. Soc.} {\bfseries 478} (2018) 3756}
  [\href{https://arxiv.org/abs/1801.00672}{{\ttfamily 1801.00672}}].

\bibitem{Liu:2022bvr}
B.~Liu and V.~Bromm, \emph{{Accelerating Early Massive Galaxy Formation with
  Primordial Black Holes}},
  \href{https://doi.org/10.3847/2041-8213/ac927f}{\emph{Astrophys. J. Lett.}
  {\bfseries 937} (2022) L30}
  [\href{https://arxiv.org/abs/2208.13178}{{\ttfamily 2208.13178}}].

\bibitem{Hutsi:2022fzw}
G.~H\"utsi, M.~Raidal, J.~Urrutia, V.~Vaskonen and H.~Veerm\"ae, \emph{{Did
  JWST observe imprints of axion miniclusters or primordial black holes?}},
  \href{https://doi.org/10.1103/PhysRevD.107.043502}{\emph{Phys. Rev. D}
  {\bfseries 107} (2023) 043502}
  [\href{https://arxiv.org/abs/2211.02651}{{\ttfamily 2211.02651}}].

\bibitem{Carr:1975qj}
B.J.~Carr, \emph{{The Primordial black hole mass spectrum}},
  \href{https://doi.org/10.1086/153853}{\emph{Astrophys. J.} {\bfseries 201}
  (1975) 1}.

\bibitem{Deng:2017uwc}
H.~Deng and A.~Vilenkin, \emph{{Primordial black hole formation by vacuum
  bubbles}}, \href{https://doi.org/10.1088/1475-7516/2017/12/044}{\emph{JCAP}
  {\bfseries 12} (2017) 044}
  [\href{https://arxiv.org/abs/1710.02865}{{\ttfamily 1710.02865}}].

\bibitem{Deng:2020mds}
H.~Deng, \emph{{Primordial black hole formation by vacuum bubbles. Part II}},
  \href{https://doi.org/10.1088/1475-7516/2020/09/023}{\emph{JCAP} {\bfseries
  09} (2020) 023} [\href{https://arxiv.org/abs/2006.11907}{{\ttfamily
  2006.11907}}].

\bibitem{Kusenko:2020pcg}
A.~Kusenko, M.~Sasaki, S.~Sugiyama, M.~Takada, V.~Takhistov and E.~Vitagliano,
  \emph{{Exploring Primordial Black Holes from the Multiverse with Optical
  Telescopes}},
  \href{https://doi.org/10.1103/PhysRevLett.125.181304}{\emph{Phys. Rev. Lett.}
  {\bfseries 125} (2020) 181304}
  [\href{https://arxiv.org/abs/2001.09160}{{\ttfamily 2001.09160}}].

\bibitem{Maeso:2021xvl}
D.N.~Maeso, L.~Marzola, M.~Raidal, V.~Vaskonen and H.~Veerm\"ae,
  \emph{{Primordial black holes from spectator field bubbles}},
  \href{https://doi.org/10.1088/1475-7516/2022/02/017}{\emph{JCAP} {\bfseries
  02} (2022) 017} [\href{https://arxiv.org/abs/2112.01505}{{\ttfamily
  2112.01505}}].

\bibitem{Hawking:1982ga}
S.W.~Hawking, I.G.~Moss and J.M.~Stewart, \emph{{Bubble Collisions in the Very
  Early Universe}}, \href{https://doi.org/10.1103/PhysRevD.26.2681}{\emph{Phys.
  Rev. D} {\bfseries 26} (1982) 2681}.

\bibitem{Kodama:1982sf}
H.~Kodama, M.~Sasaki and K.~Sato, \emph{{Abundance of Primordial Holes Produced
  by Cosmological First Order Phase Transition}},
  \href{https://doi.org/10.1143/PTP.68.1979}{\emph{Prog. Theor. Phys.}
  {\bfseries 68} (1982) 1979}.

\bibitem{Lewicki:2019gmv}
M.~Lewicki and V.~Vaskonen, \emph{{On bubble collisions in strongly supercooled
  phase transitions}},
  \href{https://doi.org/10.1016/j.dark.2020.100672}{\emph{Phys. Dark Univ.}
  {\bfseries 30} (2020) 100672}
  [\href{https://arxiv.org/abs/1912.00997}{{\ttfamily 1912.00997}}].

\bibitem{Kawana:2021tde}
K.~Kawana and K.-P.~Xie, \emph{{Primordial black holes from a cosmic phase
  transition: The collapse of Fermi-balls}},
  \href{https://doi.org/10.1016/j.physletb.2021.136791}{\emph{Phys. Lett. B}
  {\bfseries 824} (2022) 136791}
  [\href{https://arxiv.org/abs/2106.00111}{{\ttfamily 2106.00111}}].

\bibitem{Liu:2021svg}
J.~Liu, L.~Bian, R.-G.~Cai, Z.-K.~Guo and S.-J.~Wang, \emph{{Primordial black
  hole production during first-order phase transitions}},
  \href{https://doi.org/10.1103/PhysRevD.105.L021303}{\emph{Phys. Rev. D}
  {\bfseries 105} (2022) L021303}
  [\href{https://arxiv.org/abs/2106.05637}{{\ttfamily 2106.05637}}].

\bibitem{Jung:2021mku}
T.H.~Jung and T.~Okui, \emph{{Primordial black holes from bubble collisions
  during a first-order phase transition}},
  \href{https://arxiv.org/abs/2110.04271}{{\ttfamily 2110.04271}}.

\bibitem{Hashino:2022tcs}
K.~Hashino, S.~Kanemura, T.~Takahashi and M.~Tanaka, \emph{{Probing first-order
  electroweak phase transition via primordial black holes in the effective
  field theory}},
  \href{https://doi.org/10.1016/j.physletb.2023.137688}{\emph{Phys. Lett. B}
  {\bfseries 838} (2023) 137688}
  [\href{https://arxiv.org/abs/2211.16225}{{\ttfamily 2211.16225}}].

\bibitem{Ashoorioon:2020hln}
A.~Ashoorioon, A.~Rostami and J.T.~Firouzjaee, \emph{{Examining the end of
  inflation with primordial black holes mass distribution and gravitational
  waves}}, \href{https://doi.org/10.1103/PhysRevD.103.123512}{\emph{Phys. Rev.
  D} {\bfseries 103} (2021) 123512}
  [\href{https://arxiv.org/abs/2012.02817}{{\ttfamily 2012.02817}}].

\bibitem{Huang:2022him}
P.~Huang and K.-P.~Xie, \emph{{Primordial black holes from an electroweak phase
  transition}}, \href{https://doi.org/10.1103/PhysRevD.105.115033}{\emph{Phys.
  Rev. D} {\bfseries 105} (2022) 115033}
  [\href{https://arxiv.org/abs/2201.07243}{{\ttfamily 2201.07243}}].

\bibitem{Kawana:2022lba}
K.~Kawana, P.~Lu and K.-P.~Xie, \emph{{First-order phase transition and fate of
  false vacuum remnants}},
  \href{https://doi.org/10.1088/1475-7516/2022/10/030}{\emph{JCAP} {\bfseries
  10} (2022) 030} [\href{https://arxiv.org/abs/2206.09923}{{\ttfamily
  2206.09923}}].

\bibitem{Kawana:2022olo}
K.~Kawana, T.~Kim and P.~Lu, \emph{{PBH Formation from Overdensities in Delayed
  Vacuum Transitions}},  \href{https://arxiv.org/abs/2212.14037}{{\ttfamily
  2212.14037}}.

\bibitem{Coleman:1977py}
S.R.~Coleman, \emph{{The Fate of the False Vacuum. 1. Semiclassical Theory}},
  \href{https://doi.org/10.1103/PhysRevD.16.1248}{\emph{Phys. Rev. D}
  {\bfseries 15} (1977) 2929}.

\bibitem{Callan:1977pt}
C.G.~Callan, Jr. and S.R.~Coleman, \emph{{The Fate of the False Vacuum. 2.
  First Quantum Corrections}},
  \href{https://doi.org/10.1103/PhysRevD.16.1762}{\emph{Phys. Rev. D}
  {\bfseries 16} (1977) 1762}.

\bibitem{Linde:1981zj}
A.D.~Linde, \emph{{Decay of the False Vacuum at Finite Temperature}},
  \href{https://doi.org/10.1016/0550-3213(83)90072-X}{\emph{Nucl. Phys. B}
  {\bfseries 216} (1983) 421}.

\bibitem{Bodeker:2009qy}
D.~Bodeker and G.D.~Moore, \emph{{Can electroweak bubble walls run away?}},
  \href{https://doi.org/10.1088/1475-7516/2009/05/009}{\emph{JCAP} {\bfseries
  05} (2009) 009} [\href{https://arxiv.org/abs/0903.4099}{{\ttfamily
  0903.4099}}].

\bibitem{Bodeker:2017cim}
D.~Bodeker and G.D.~Moore, \emph{{Electroweak Bubble Wall Speed Limit}},
  \href{https://doi.org/10.1088/1475-7516/2017/05/025}{\emph{JCAP} {\bfseries
  05} (2017) 025} [\href{https://arxiv.org/abs/1703.08215}{{\ttfamily
  1703.08215}}].

\bibitem{Hoche:2020ysm}
S.~H\"oche, J.~Kozaczuk, A.J.~Long, J.~Turner and Y.~Wang, \emph{{Towards an
  all-orders calculation of the electroweak bubble wall velocity}},
  \href{https://doi.org/10.1088/1475-7516/2021/03/009}{\emph{JCAP} {\bfseries
  03} (2021) 009} [\href{https://arxiv.org/abs/2007.10343}{{\ttfamily
  2007.10343}}].

\bibitem{Gouttenoire:2021kjv}
Y.~Gouttenoire, R.~Jinno and F.~Sala, \emph{{Friction pressure on relativistic
  bubble walls}}, \href{https://doi.org/10.1007/JHEP05(2022)004}{\emph{JHEP}
  {\bfseries 05} (2022) 004}
  [\href{https://arxiv.org/abs/2112.07686}{{\ttfamily 2112.07686}}].

\bibitem{Sagunski:2023ynd}
L.~Sagunski, P.~Schicho and D.~Schmitt, \emph{{Supercool exit: Gravitational
  waves from QCD-triggered conformal symmetry breaking}},
  \href{https://arxiv.org/abs/2303.02450}{{\ttfamily 2303.02450}}.

\bibitem{Lewicki:2022nba}
M.~Lewicki, V.~Vaskonen and H.~Veerm\"ae, \emph{{Bubble dynamics in fluids with
  N-body simulations}},
  \href{https://doi.org/10.1103/PhysRevD.106.103501}{\emph{Phys. Rev. D}
  {\bfseries 106} (2022) 103501}
  [\href{https://arxiv.org/abs/2205.05667}{{\ttfamily 2205.05667}}].

\bibitem{Ellis:2020nnr}
J.~Ellis, M.~Lewicki and V.~Vaskonen, \emph{{Updated predictions for
  gravitational waves produced in a strongly supercooled phase transition}},
  \href{https://doi.org/10.1088/1475-7516/2020/11/020}{\emph{JCAP} {\bfseries
  11} (2020) 020} [\href{https://arxiv.org/abs/2007.15586}{{\ttfamily
  2007.15586}}].

\bibitem{Guth:1982pn}
A.H.~Guth and E.J.~Weinberg, \emph{{Could the Universe Have Recovered from a
  Slow First Order Phase Transition?}},
  \href{https://doi.org/10.1016/0550-3213(83)90307-3}{\emph{Nucl. Phys. B}
  {\bfseries 212} (1983) 321}.

\bibitem{Ellis:2018mja}
J.~Ellis, M.~Lewicki and J.M.~No, \emph{{On the Maximal Strength of a
  First-Order Electroweak Phase Transition and its Gravitational Wave Signal}},
  \href{https://doi.org/10.1088/1475-7516/2019/04/003}{\emph{JCAP} {\bfseries
  04} (2019) 003} [\href{https://arxiv.org/abs/1809.08242}{{\ttfamily
  1809.08242}}].

\bibitem{Ellis:2019oqb}
J.~Ellis, M.~Lewicki, J.M.~No and V.~Vaskonen, \emph{{Gravitational wave energy
  budget in strongly supercooled phase transitions}},
  \href{https://doi.org/10.1088/1475-7516/2019/06/024}{\emph{JCAP} {\bfseries
  06} (2019) 024} [\href{https://arxiv.org/abs/1903.09642}{{\ttfamily
  1903.09642}}].

\bibitem{Turner:1992tz}
M.S.~Turner, E.J.~Weinberg and L.M.~Widrow, \emph{{Bubble nucleation in first
  order inflation and other cosmological phase transitions}},
  \href{https://doi.org/10.1103/PhysRevD.46.2384}{\emph{Phys. Rev. D}
  {\bfseries 46} (1992) 2384}.

\bibitem{Saini:2017tsz}
A.~Saini and D.~Stojkovic, \emph{{Modified hoop conjecture in expanding
  spacetimes and primordial black hole production in FRW universe}},
  \href{https://doi.org/10.1088/1475-7516/2018/05/071}{\emph{JCAP} {\bfseries
  05} (2018) 071} [\href{https://arxiv.org/abs/1711.06732}{{\ttfamily
  1711.06732}}].

\bibitem{deJong:2021bbo}
E.~de~Jong, J.C.~Aurrekoetxea and E.A.~Lim, \emph{{Primordial black hole
  formation with full numerical relativity}},
  \href{https://doi.org/10.1088/1475-7516/2022/03/029}{\emph{JCAP} {\bfseries
  03} (2022) 029} [\href{https://arxiv.org/abs/2109.04896}{{\ttfamily
  2109.04896}}].

\bibitem{Lewicki:2023mik}
M.~Lewicki, K.~M\"u\"ursepp, J.~Pata, M.~Vasar, V.~Vaskonen and H.~Veerm\"ae,
  \emph{{Dynamics of false vacuum bubbles with trapped particles}},
  \href{https://arxiv.org/abs/2305.07702}{{\ttfamily 2305.07702}}.

\bibitem{Gouttenoire:2023naa}
Y.~Gouttenoire and T.~Volansky, \emph{{Primordial Black Holes from Supercooled
  Phase Transitions}},  \href{https://arxiv.org/abs/2305.04942}{{\ttfamily
  2305.04942}}.

\end{thebibliography}\endgroup

\end{document}